\documentclass[pre,superscriptaddress,showpacs,longbibliography,reprint]{revtex4-2} 

\usepackage[colorlinks=true,linkcolor=blue,urlcolor=blue,citecolor=blue,pdfusetitle]{hyperref}
\usepackage{color}
\usepackage[utf8]{inputenc}
\usepackage[usenames,dvipsnames]{xcolor}
\usepackage{amsmath}
\usepackage{amssymb}
\usepackage{graphicx}
\graphicspath{{graphics/}}
\usepackage{epsfig}
\usepackage{dcolumn}
\usepackage{bm}
\usepackage{mathrsfs}
\usepackage{multirow}
\usepackage[all]{xy}
\usepackage{pbox}
\usepackage{lipsum}
\usepackage{verbatim}
\usepackage{braket}
\usepackage{dsfont}
\usepackage{array}
\usepackage{makecell}
\usepackage{tabularx}
\usepackage{isotope}
\usepackage{xr}
\usepackage{float}

\begin{document}
\title{Universal and nonuniversal probability laws  in Markovian open quantum dynamics subject to generalized reset processes}
\author{Federico Carollo}
\affiliation{%
Institut f\"ur Theoretische Physik, Universit\"at T\"ubingen, Auf der Morgenstelle 14, 72076 T\"ubingen, Germany
}%
\author{Igor Lesanovsky}
\affiliation{%
Institut f\"ur Theoretische Physik, Universit\"at T\"ubingen, Auf der Morgenstelle 14, 72076 T\"ubingen, Germany
}%
\affiliation{School of Physics and Astronomy,  University of Nottingham, Nottingham, NG7 2RD, United Kingdom}
\affiliation{Centre for the Mathematics and Theoretical Physics of Quantum Non-Equilibrium Systems, University of Nottingham, Nottingham, NG7 2RD, United Kingdom}
\author{Juan P. Garrahan}
\affiliation{School of Physics and Astronomy,  University of Nottingham, Nottingham, NG7 2RD, United Kingdom}
\affiliation{Centre for the Mathematics and Theoretical Physics of Quantum Non-Equilibrium Systems, University of Nottingham, Nottingham, NG7 2RD, United Kingdom}

\begin{abstract}
We consider quantum jump trajectories of Markovian open quantum systems subject to stochastic in time resets of their state to an initial configuration. The reset events provide a partitioning of quantum trajectories into consecutive time intervals, defining sequences of random variables from the values of a trajectory observable within each of the intervals. For observables related to functions of the quantum state, we show that the probability of certain orderings in the sequences obeys a universal law. This law does not depend on the chosen observable and, in case of Poissonian reset processes, not even on the details of the dynamics. When considering (discrete) observables associated with the counting of quantum jumps, the probabilities in general lose their universal character. Universality is only recovered in cases when the probability of observing equal outcomes in a same sequence is vanishingly small, which we can achieve in a weak reset rate limit. Our results extend previous findings on classical stochastic processes [N.~R.~Smith et al., EPL {\bf 142}, 51002 (2023)] to the quantum domain and to state-dependent reset processes, shedding light on relevant aspects for the emergence of universal probability laws.
\end{abstract}

\maketitle

\section{Introduction} 
The dynamics of quantum systems which are in contact with their surroundings, usually an infinitely large thermal bath, is characterized by dissipation and stochastic effects \cite{breuer2002theory,gardiner2004}. By assuming a weak system-bath coupling and considering standard Markovian approximations, the time evolution of the average state of these open quantum systems is described by quantum master equations \cite{breuer2002theory,gardiner2004,lindblad1976,gorini1976}. The latter are implemented by Lindblad generators \cite{lindblad1976,gorini1976} and provide a complete description of the system evolution, whenever the system-environment  interaction is not monitored \cite{breuer2002theory}. On the other hand, when the interaction is monitored \cite{wiseman2009}, for instance by means of a detector counting the quanta exchanged between the system and the environment, quantum master equations provide a description of the system evolution averaged over all possible realizations of the system-environment interaction. Due to the stochastic nature of the emission and of the absorption of energy quanta \cite{gardiner2004,wiseman2009}, single realizations of the dynamics can only be described by quantum stochastic processes, generating quantum trajectories \cite{belavkin1989,belavkin1989b,plenio1998,barchielli1991}. For counting experiments, quantum jump trajectories provide the appropriate unravelling of quantum master equations \cite{dalibard1992,dum1992,gardiner1992} into single dynamical runs. In one such trajectory, the state of the system undergoes a continuous evolution, conditional on not having detected any event, interrupted by abrupt changes, or jumps, of the system state. The latter  take place at random times and are associated with the exchange of energy quanta  \cite{nagourney1986,sauter1986,bergquist1986}. 

\begin{figure}[t]
\centering
\includegraphics[width=0.8\columnwidth]{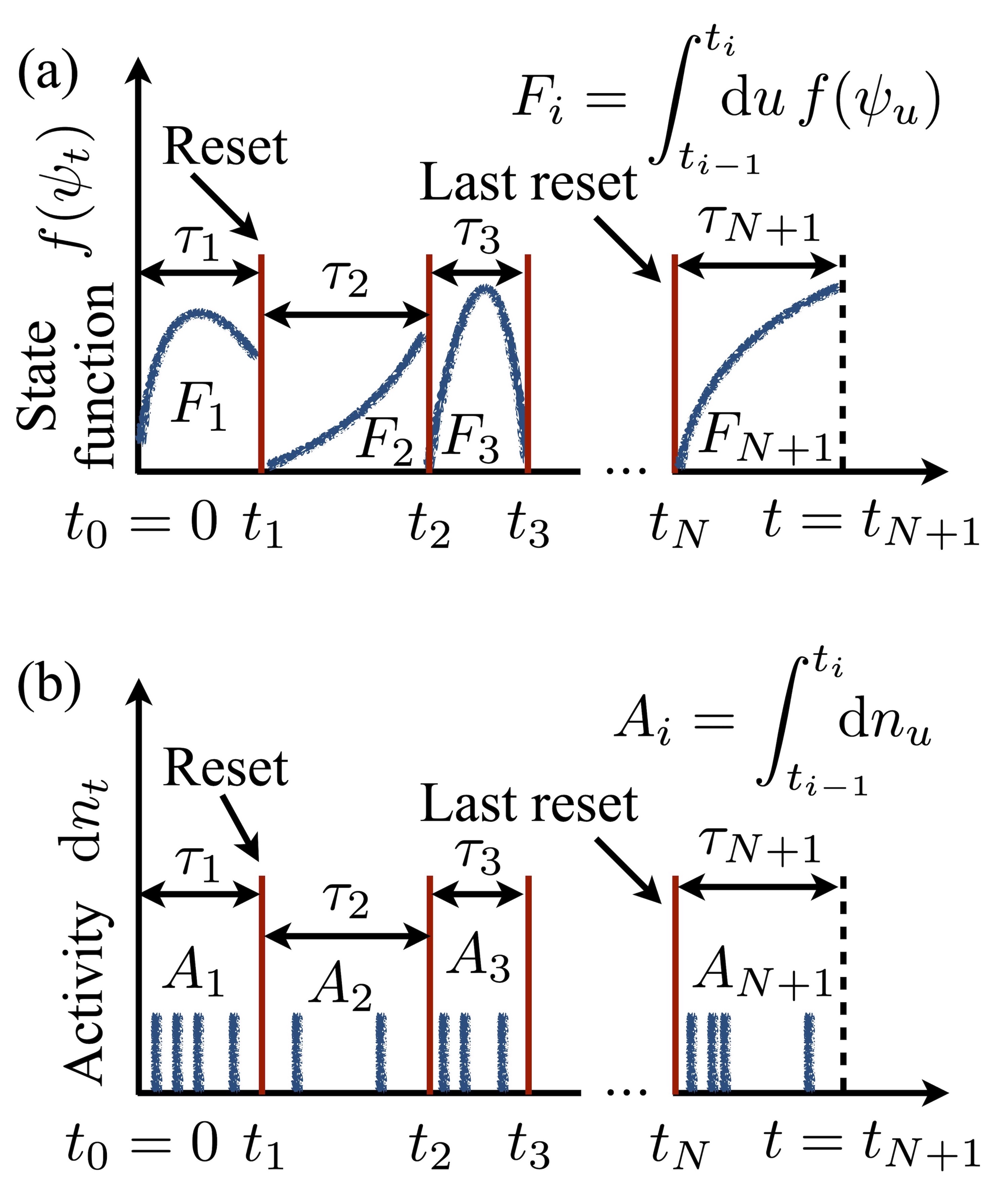}
\caption{{\bf Open quantum reset dynamics and trajectory observables.} (a) A quantum trajectory of an open quantum system subject to stochastic resetting is naturally partitioned into different time-windows, with extension $\tau_i=[t_{i-1},t_i]$,  by the times $t_i$ in which reset events occur. The last time-window is special since it does not end with a reset event. Given a function of the state $f(\psi)$, it is possible to construct random variables within the different time-windows, as $F_i=\int_{t_{i-1}}^{t_i} {\rm d} u f(\psi_u)$. As we show, the probability of observing certain sequences for the observables $F_i$ are universal, i.e., they do not depend on the open quantum dynamics and not even on the chosen function $f(\psi)$. (b) When considering instead of a state function the number of jump events (or dynamical activity)  observed within the different time-windows $A_i$, the probability of the same sequences are not universal in general. This due to the discrete nature of the random variable $A_i$.}
\label{Fig1}
\end{figure} 

In the last few years there has been growing interest in exploring how both classical and quantum stochastic dynamics are affected by the presence of reset processes \cite{evans2020r,gupta2022,pal2022}. These consist of random-in-time re-initializations of the state of the system back to an initial configuration or state. In the simplest scenario, the random times between reset events are distributed exponentially (Poissonian reset process) and  the reset rate does not depend on the instantaneous state of the system \cite{evans2011}. Reset processes have been extensively studied in classical systems, where they have been shown to give rise to nonequilibrium stationary states \cite{evans2011,evans2014,majumdar2015,pal2015,mendez2016,maes2017,falcon2017,fuchs2016,nagar2016,magoni2020,biroli2023} as well as  to improve the efficiency of search processes \cite{kusmierz2014,campos2015,pal2017,bressloff2020,besga2020,faisant2021,debruyne2022}. 
In quantum systems, reset processes have been studied focussing on the spectral properties of the dynamical generator \cite{rose2018}, on the emergent stationary behavior
\cite{mukherjee2018,perfetto2021,haack2021,magoni2022,sevilla2023}, on their impact on hitting times \cite{yin2023,yin2023b} and on how they affect the probability distribution of emission events \cite{perfetto2022} compared to 
the classical case \cite{meylahn2015,harris2017,hollander2019,coghi2020,monthus2021}.

When an open quantum system is subject to $N$ stochastic resettings, its quantum trajectories can be divided into $N+1$ consecutive time-intervals \cite{perfetto2022}, delimited by the times at which the reset events occur. This is sketched in Fig.~\ref{Fig1}. Given a function of the state, $f(\psi)$, such as a measure of  coherence or entanglement, its time-integral in each interval gives rise to a sequence of random variables, $\{F_i\}_{i=1}^{N+1}$, see Fig.~\ref{Fig1}(a).  Interestingly, for classical stochastic processes it was shown 
in Ref.~\cite{smith2023}
that the presence of Poissonian resets gives rise to universal probability laws for the elements of the sequences $\{F_i\}$ to obey certain relations. Consider the example of the 
probability that $F_1>F_i$, $\forall i$, i.e., the first element being the largest in the sequence \cite{smith2023}. The universal character of the probability of such an occurrence is due to the fact that the reset process renders the elements of the sequence statistically independent and identically distributed. Consequently, there is no preference on which element should be the largest, 
and the probability is given by the inverse of the number elements in the sequence, irrespective of the chosen observable and of the details of the dynamics \cite{smith2023}. 

In this paper, we demonstrate that analogous universal probability laws to those of Ref.~\cite{smith2023} can hold for Markovian open quantum systems but only when considering certain trajectory observables. We consider generalized reset processes with rates that can in general depend on the instantaneous state of the system, making the reset times non-Poissonian. In this situation, we find that for trajectory observables which are time-integrated functions of the state, a universal character of the probability of a given ordering of the sequence $\{F_i\}$ exists
as long as the last time-interval is disregarded, see Fig.~\ref{Fig1}(a): this is a consequence of the non-Poissonian nature of the generalized reset process and also applies to classical systems with non-Poissonian reset clocks, as recently observed in Ref.~\cite{godreche2023}. 
Moreover, there is an inevitable interplay between the generalized reset process and the intrinsic open quantum dynamics of the system so that probabilities are universal solely in the sense that they do not depend on the chosen observable but can actually depend on the specific dynamics considered. 

We also consider a class of (discrete) observables, constructed from the number of jump events occurred in the different time-intervals,  as sketched in Fig.~\ref{Fig1}(b). In this case, we do not find any universal behavior for the above-mentioned probabilities. Instead, they become dependent on the observable and on the details of the dynamics, for both Poissonian and non-Poissonian reset processes. This breakdown of universality is due to the fact that when observables assume a discrete set of values there is in general a non-zero probability that two outcomes in the sequence are strictly equal (see also Ref.~\cite{godreche2023}). This makes the probability of different orderings dependent on the details of the probability distribution of the random variable within a single time-interval. As we show, in this case universal laws can be recovered by considering a weak reset-rate limit. Here, the number of jumps still assumes a discrete set of values but the probability for two outcomes in the sequence being equal is vanishingly small in the length of the time-intervals, which get longer and longer the weaker the reset rate. 
 
Our results show that the mere presence of reset events is not sufficient to observe universal probability laws. Universality indeed breaks down when considering discrete-valued trajectory observables characterized by a finite probability of finding two equal outcomes in the same sequence. Our findings can be immediately generalized to include classical continuous-time Markov chain dynamics in the presence of stochastic resetting, since the latter can also be encoded within a Lindblad formalism. 

The rest of the paper is organized as follows. Section II describes the standard formalism of Markovian open quantum dynamics and stochastic trajectories, while Sec.~III describes the class of generalized stochastic resetting problems we study. Section IV considers the existence of universal probability laws for trajectory observables which are functions of the state. Section V considers the corresponding problem for quantum jump observables. Section VI rationalizes our findings by studying a simple model that captures the essential physics. In Sec.~VII we provide a discussion and the conclusions.

\section{Markovian open quantum dynamics and quantum trajectories}

\subsection{Markovian Lindblad dynamics}
Any (finite-dimensional) quantum system can be associated with a separable Hilbert space $\mathcal{H}$, having suitable dimension $d$. This space contains all possible pure states of the system. It is spanned by the orthonormal basis vectors $\{\ket{m}\}_{m=1}^d$, obeying the conditions $\braket{m|n}=\delta_{mn}$. Statistical mixtures of pure states can be encoded through mixed-state density matrices, $\rho=\sum_i p(\psi_i) \ket{\psi_i}\!\bra{\psi_i}$, for some well-defined probability $p(\psi_i)$ over pure states. 

A description of quantum systems in terms of density matrices is convenient whenever considering open quantum time evolutions. In the simplest case of Markovian open quantum dynamics, the evolution of the density matrix is implemented by a quantum master equation $\dot{\rho}_t=\mathcal{L}[\rho_t]$ \cite{breuer2002theory}, with  Lindblad generator  \cite{lindblad1976,gorini1976}
\begin{equation}
    \mathcal{L}[\rho]=-i[H,\rho]+\sum_{\alpha=1}^{D}\gamma_\alpha\left( J_\alpha \rho J_\alpha^\dagger -\frac{1}{2}\left\{\rho ,J^\dagger_\alpha J_\alpha\right\}\right)\, .
    \label{lindblad}
\end{equation}
The operator $H=H^\dagger$ represents the Hamiltonian of the system, while the operators $J_\alpha$ are the so-called jump operators, each of the $D$ different ones  associated with the rate $\gamma_\alpha$. The latter operators encode how the dynamics of the system is affected by the presence of an external  environment. For instance, in quantum optics, jump operators are connected with the emission  (absorption) of photons into (from) the environment and describe how the quantum state changes when these events occur. 
For later convenience, we decompose the (probability conserving) Lindblad generator in terms of the (non-probability conserving) super-operators
$$
\mathcal{J}_\alpha[\rho]   
    =\gamma_\alpha J_\alpha \rho J_\alpha^\dagger\, ,\quad 
\mathcal{L}_\infty[\rho]
    =-iH_{\rm eff}\rho +i\rho H_{\rm eff}^\dagger \, , 
$$
where 
$$
H_{\rm eff}=H-\frac{i}{2}\sum_\alpha \gamma_\alpha J^\dagger_\alpha J_\alpha \, , 
$$
and $\mathcal{L} =  \mathcal{L}_\infty + \sum_\alpha \mathcal{J}_\alpha$.
The dynamics governed by the generator $\mathcal{L}$, Eq.~\eqref{lindblad}, is nonunitary and deterministic. It can be interpreted as the dynamics describing the state of the system averaged over all possible realizations of the interaction between the system and the environment, which may for instance be monitored in experiments \cite{wiseman2009}. Single dynamical realizations, or quantum trajectories, of the open system are instead captured by quantum stochastic processes \cite{belavkin1989,belavkin1989b,plenio1998,barchielli1991,dalibard1992,dum1992,gardiner1992}. In the following, we consider the situation in which the average dynamics in Eq.~\eqref{lindblad} is unravelled into quantum jump trajectories \cite{barchielli1991,dalibard1992,dum1992,gardiner1992,plenio1998}.

\subsection{Quantum jump trajectories and their probabilities}
Modern experiments with quantum systems allow for the  monitoring of the system-environment interaction through the detection of the quanta exchanged between them, such as, for instance, the photons emitted or absorbed by the system. Since such continuous monitoring is in fact a proper measurement process on the composite system \cite{wiseman2009,gardiner2004}, the dynamics in single (experimental) realizations is stochastic and, in the case of the counting processes of interest in this work, it can be described by means of so-called quantum jump trajectories \cite{dalibard1992,dum1992,gardiner1992}. 

In a single realization of the quantum jump process described, on average, by the dynamics in Eq.~\eqref{lindblad}, the quantum state of the system evolves according to the (nonlinear) quantum stochastic equation \cite{belavkin1989,plenio1998} 
\begin{equation}
\begin{split}
{\rm d}\psi_{t} &= \left(\frac{e^{{\rm d }t\mathcal{L}_\infty}[\psi_t]}{{\rm Tr}\left(e^{{\rm d }t\mathcal{L}_\infty}[\psi_t]\right)}-\psi_t\right)\left(1-\sum_{\alpha}{\rm d} n^{\alpha}_t\right) \\
&+\sum_{\alpha}\left(\frac{\mathcal{J}_\alpha[\psi_t]}{{\rm Tr}\left(\mathcal{J}_\alpha [\psi_t]\right)}-\psi_t\right){\rm d}n_t^\alpha\, .
\end{split}
\label{eq:stoc}
\end{equation}
Here, the state $\psi_t$ is the state of the system at time $t$, while ${\rm d}\psi_t:=\psi_{t+{\rm d}t}-\psi_t$ represents its increment in the infinitesimal time step ${\rm d}t$. The quantities ${\rm d}n_t^\alpha$ are the proper random variables, which can assume either the value $0$ or the value $1$. The probability for each of the noises to be $1$, when the system is at time $t$ in state $\psi_t=\psi$, is given by 
\begin{equation}
P({\rm d}n_t^\alpha=1|\psi_t=\psi)= {\rm Tr} \left(\mathcal{J}_\alpha[\psi]\right) {\rm d} t\, .
\label{p_event}
\end{equation}
Since this probability is of the order of the time increment ${\rm d}t$, effectively at most only one noise ${\rm d }n_t^\alpha$ can be different from zero at each time $t$. When one noise is equal to one, let us say ${\rm d }n_t^{\bar{\alpha}}=1$, the updated state is obtained via the map $\mathcal{J}_{\bar{\alpha}}$ implementing an abrupt change of the state of the system. From the viewpoint of an experiment this situation corresponds to the detection of an event associated with the jump operator $J_{\bar{\alpha}}$. When, instead, ${\rm d }n_t^\alpha=0$, $\forall \alpha$, the state evolves continuously through the ``no-jump" dynamical map $e^{{\rm d}t\mathcal{L}_\infty}$. This corresponds to the detector signalling absence of emission or of absorption events at time $t$. 

A quantum jump trajectory thus consists of the whole history of the system state $\{\psi_u\}_{u\in[0,t]}$, from the initial time to the final one. It is completely specified by the initial state together with the times and the types of the occurred jumps. We denote quantum trajectories with the symbol $\omega_{\vec{t},\vec{\alpha}}^{\psi_0}=\{\psi_u\}_{u\in[0,t]}$. Here, $\psi_0$ denotes the initial state of the system, the vector $\vec{t}=(0,v_{\alpha_1},v_{\alpha_2},\dots v_{\alpha_m},t)$ specifies the initial and the final time as well as the times $v_{\alpha_i}$ in which the jumps took place, while $\vec{\alpha}=(\alpha_1,\alpha_2,\dots \alpha_m)$ indicates the type of the $m$ occurred jumps. The state of the system at the final time $t$ can be recovered by reconstructing the history of the evolution by combining the maps which sequentially acted on the initial state. In this regard, we can define the unnormalized state $\psi_{\vec{t},\vec{\alpha}}^{\rm u}=\Lambda_{\vec{t},\vec{\alpha}}[\psi_0]$, where 
\begin{equation}
\begin{split}
\Lambda_{\vec{t},\vec{\alpha}}[\cdot]=&e^{(t-v_{\alpha_m})\mathcal{L}_\infty}\circ\mathcal{J}_{\alpha_{m}}\circ\dots     \\
&\dots\mathcal{J}_{\alpha_2}\circ e^{(v_{\alpha_2}-v_{\alpha_1})\mathcal{L}_\infty}\circ \mathcal{J}_{\alpha_1}\circ e^{v_{\alpha_1}{\mathcal{L}}_\infty}[\cdot]
\end{split}
\label{unnor_state}
\end{equation}
is associated with a trajectory in which jump $\alpha_1$ occurred at time $v_{\alpha_1}$, jump $\alpha_2$ at time $v_{\alpha_2}$, and so on up to a last jump $\alpha_m$ at time $v_{\alpha_m}$. Importantly, the final state of such trajectory is the normalized state $\psi_{\vec{t},\vec{\alpha}}=\psi_{\vec{t},\vec{\alpha}}^{\rm u}/{\rm Tr}\left(\psi_{\vec{t},\vec{\alpha}}^{\rm u}\right)$ and the probability (density function) of such a trajectory is given by 
\begin{equation}
\label{prob_trajectory}
P\left(\omega_{\vec{t},\vec{\alpha}}^{\psi_0}\right)={\rm Tr}\left(\psi_{\vec{t},\vec{\alpha}}^{\rm u}\right)={\rm Tr}\left(\Lambda_{\vec{t},\vec{\alpha}}[\psi_0]\right)\, .
\end{equation}

\subsection{Observables of quantum trajectories}
An observable of a quantum trajectory is any possible function of the whole history of the state $\omega_{\vec{t},\vec{\alpha}}^{\psi_0}$. While these functions can be in principle very general, the focus is usually on two important classes of additive (or time-integrated) observables \cite{lecomte2007,garrahan2009,garrahan2010,jack2010,carollo2019,carollo2021}. The first class is the one consisting of  time-integrated functions of the state. Considering any linear or nonlinear function $f(\psi)$, the latter are defined as 
\begin{equation}
F\left(\omega_{\vec{t},\vec{\alpha}}^{\psi_0}\right)=\int_0^t {\rm d}u \,  f(\psi_u)\, .
\label{time_int_state}
\end{equation}
Through the probability of trajectories introduced in Eq.~\eqref{prob_trajectory}, it is possible to write the probability density function of observing a value $F=\bar{F}$ up to a time $t$ as 
\begin{equation}
P(F=\bar{F}|t)=\sum_{\forall \vec{t},\vec{\alpha}}\delta\left[F\left(\omega_{\vec{t},\vec{\alpha}}^{\psi_0}\right)-\bar{F}\right]P\left(\omega_{\vec{t},\vec{\alpha}}^{\psi_0}\right)\, .
\label{prob-F-t}
\end{equation}
Here, the sum is over all trajectories and thus includes any possible number of jumps, an integral over all possible jump times and sums over all possible jump types.
The normalization of this probability function follows from 
\begin{equation}
\int {\rm d}F \, P(F|t)=\sum_{\forall \vec{t},\vec{\alpha}}P\left(\omega_{\vec{t},\vec{\alpha}}^{\psi_0}\right)={\rm Tr}\left(e^{t\mathcal{L}}[\psi_0]\right)=1\, .
\label{normaliz}
\end{equation}
The last equality stems from the trace-preservation of $\mathcal{L}$ while the second to last equality encodes the fact that the ensemble of trajectories provides an unravelling of the Lindblad dynamics. Mathematically, this can be seen by expanding the propagator $e^{t\mathcal{L}}$ in Dyson series, by considering the map $\sum_\alpha\mathcal{J}_\alpha$ as an ``interaction". 
To conclude this section, we note that the formalism introduced here is valid for both pure and mixed initial states $\psi_0$. For the sake of clarity, however, in the following we assume an initial pure state, i.e., $\psi_0=\ket{\psi_0}\!\bra{\psi_0}$.

\section{Open quantum dynamics subject to state-dependent resets}

We now consider the case in which the dynamics implemented by the Lindblad generator $\mathcal{L}$ is interspersed with reset events. The basic idea is that the open quantum stochastic process runs up to a random time $\tau$, after which the state of the system is projected back into the initial state $\psi_0$. After that, the stochastic dynamics starts over again. In the simplest case of  a Poissonian reset process, the reset rate does not depend on the instantaneous state of the system and the time interval $\tau $ between reset events is distributed exponentially. That is, the probability density function for the reset times is 
$$
 p(\tau)=\Gamma  e^{-\Gamma \tau }\, ,
$$
with $\Gamma$ being the reset rate. The addition of such a reset process on top of the Lindblad dynamics leads to a new open quantum dynamics described by the Lindblad generator \cite{evans2011,rose2018}
\begin{equation}
    \tilde{\mathcal{L}}[\rho]=\mathcal{L}[\rho]+\Gamma \left(\psi_0 - \rho\right)\, .
    \label{Lind-Poisson-reset}
\end{equation}
In what follows, we shall consider the more general case in which the reset rate depends on the instantaneous state of the quantum system. 

\subsection{Generalized reset process}
In order to implement a (Markovian) reset process with rate depending on the specific basis state $\ket{m}$, we introduce the map
\begin{equation}
\mathcal{W}[\rho]:=\sum_{m=1}^d \Gamma_m \ket{\psi_0}\!\bra{m}\rho \ket{m}\!\bra{\psi_0}\, ,
\label{reset_jump_operators}
\end{equation}
as well as the associated ``no-reset" map 
$$
\mathcal{R}[\rho]=\frac{1}{2}\left\{\sum_{m=1}^d \Gamma_m \ket{m}\!\bra{m} ,\rho\right\}\, . 
$$
The map in Eq.~\eqref{reset_jump_operators}  allows for a clear identification of the jump operators, $\ket{\psi_0}\!\bra{m}$, implementing the reset process. The rates $\Gamma_m$ take into account that the rate of resetting the state to $\psi_0$ depends on the configuration $\ket{m}$. Due to the presence of  different reset rates, the sum appearing inside the anti-commutator in the no-reset term $\mathcal{R}$ is not proportional to the identity, so that the distribution of the reset times is in general not Poissonian. 
The open quantum dynamics of the system subject to our generalized reset process, encoded in the maps $\mathcal{W}$ and $\mathcal{R}$, is described by a Lindblad quantum master equation with generator 
\begin{equation}
\tilde{\mathcal{L}}[\rho]=\mathcal{L}[\rho]+\mathcal{W}[\rho]-\mathcal{R}[\rho]\, .
\label{Lind-Reset}
\end{equation}
When $\Gamma_m=\Gamma$ for all $m=1,2,\dots d$, the above equation reduces to the Poisson reset case, Eq.~\eqref{Lind-Poisson-reset}.

\subsection{Quantum trajectories with reset events}
In this section,  we will study the structure of quantum trajectories in the presence of the reset process, i.e., of trajectories resulting from the Lindblad generator of Eq.~\eqref{Lind-Reset}. Recalling the interpretation of Eq.~\eqref{eq:stoc}, the quantum stochastic process considered is such that, if no jumps and no reset events occur at a given time, the system evolves according to the generator
$$
\tilde{\mathcal{L}}_\infty[\rho]=\mathcal{L}_\infty[\rho] -\mathcal{R}[\rho]\,.
$$
If instead an emission occurs, then the state changes through the application of the corresponding jump operator, as it was discussed in the context of Eq.~\eqref{eq:stoc}. On the other hand, when a reset event takes place, the state of the system is brought back to $\psi_0$. 

Within a time-window between two reset events, each trajectory can still be characterized by an overall continuous dynamics, however now generated by $\tilde{\mathcal{L}}_\infty$, interspersed at random times by stochastic jump events. As before, we can introduce the map 
\begin{equation}
\begin{split}
\tilde{\Lambda}_{\vec{\tau},\vec{\alpha}}[\cdot]=&e^{(\tau-v_{\alpha_m})\tilde{\mathcal{L}}_\infty}\circ\mathcal{J}_{\alpha_{m}}\circ\dots     \\
&\dots\mathcal{J}_{\alpha_2}\circ e^{(v_{\alpha_2}-v_{\alpha_1})\tilde{\mathcal{L}}_\infty}\circ \mathcal{J}_{\alpha_1}\circ e^{v_{\alpha_1}\tilde{\mathcal{L}}_\infty}[\cdot]\, ,
\end{split}
\label{unnor_state_reset}
\end{equation}
and define the unnormalized state $\tilde{\psi}_{\vec{\tau},\vec{\alpha}}^{\rm u}$
whose trace yields the probability of observing the trajectory $\omega_{\vec{\tau},\vec{\alpha}}^{\psi_0}$ between two reset events (note indeed that no jump operator associated with the reset appears in the above equation). From these probabilities, we can also compute the probability of observing a given outcome for the observable $F$ within two reset events, along the lines that lead to Eq.~\eqref{prob-F-t}.

\subsection{Probability of trajectory observables in the presence of resets }
We now proceed with characterizing the open quantum reset dynamics. Within a single  trajectory one can observe different reset events, which partition the trajectory into several time-intervals, each one associated with a different value of the observable $F$. In particular, we denote by $F_1$ the observable in the time-interval delimited by the initial time and the time of the first reset event. $F_2$ is then the value of the observable in the time-interval delimited by the time of the first reset event and that of the second one and so on [cf.~Fig.~\ref{Fig1}(a)]. The last time-interval is special, since it does not terminate with a reset event but rather with the final observation time of the quantum trajectory. For this reason, if the trajectory is characterized by $N$ reset events, there will be $N+1$ time-intervals and thus $N+1$ random variables $F_1,F_2,\dots F_{N+1}$, as illustrated in Fig.~\ref{Fig1}(a). 

The probability density function of observing $N$ reset events, specifically occurring at the times $\{t_i\}_{i=1}^N$, and the outcome $\{F_i\}_{i=1}^{N+1}$ for the random variables up to a final time $t$ is given by (we set $\tau_k=t_{k}-t_{k-1}$, with $t_0=0$ and $t_{N+1}=t$)
\begin{equation}
\begin{split}
P(\{F_i\},&\{ t_i\},N|t)=\delta \left(t-\sum_{i=1}^{N+1} \tau_i \right)\times \\
&\times {\rm Tr}\left( \tilde{\Lambda}^{F_{N+1}}_{\tau_{N+1}}[\psi_0]\right) \prod_{i=1}^{N}{\rm Tr}\left(\mathcal{W}\circ{\tilde{\Lambda}}^{F_i}_{\tau_i}[\psi_0]\right)\, .
\end{split}
\label{prob_traj}
\end{equation}
In the above equation, we have introduced the map 
\begin{equation}
    \tilde{\Lambda}_\tau^{\bar{F}}[\psi_0]:=\sum_{\forall \vec{\tau},\vec{\alpha}}\delta(F(\omega_{\vec{\tau},\vec{\alpha}}^{\psi_0})-{\bar{F}})\tilde{\Lambda}_{\vec{\tau},\vec{\alpha}}[\psi_0]\, .
\end{equation}
This map encodes the sum over all possible trajectories, which start from $\psi_0$ and are free of reset events up to time $\tau$ and for which the chosen observable assumes the value $\bar{F}$.  The factorization of the probability in Eq.~\eqref{prob_traj} is a consequence of the fact that the reset map $\mathcal{W}$ always reinitializes the system in $\psi_0$, making time-intervals between different reset events independent. Furthermore, using the maps $\tilde{\Lambda}_\tau^{\bar{F}}$ makes it possible to write the probability of observing the value $\bar F$, in a trajectory free of reset events and starting from $\psi_0$, as ${\rm Tr}(\tilde{\Lambda}_\tau^{\bar{F}}[\psi_0])$. If the same trajectory terminates with a reset event, the probability of observing  $\bar{F}$ is given by ${\rm Tr}(\mathcal{W}\circ \tilde{\Lambda}_\tau^{\bar{F}}[\psi_0])$. 

By integrating the probability in Eq.~\eqref{prob_traj} over all possible times in which the reset events can occur, we find the probability of observing $N$ reset events associated with the sequence $\{F_i\}$ for the random variables in the different time-intervals as 
\begin{equation}
\begin{split}
P(\{F_i\},N|t)&=\int_0^t {\rm d}t_{N}\int_0^{t_{N}}{\rm d }t_{N-1}\dots \int_0^{t_2}{\rm d}t_1 \times \\
&\times {\rm Tr}\left( \tilde{\Lambda}^{F_{N+1}}_{\tau_{N+1}}[\psi_0]\right) \prod_{i=1}^{N}{\rm Tr}\left(\mathcal{W}\circ\tilde{\Lambda}^{F_{i}}_{\tau_i}[\psi_0]\right)\, .
\label{p_seq_F}
\end{split}
\end{equation}
In what follows, we exploit the product structure and the similarity between all the factors in the integrand to derive general results on the probability of observing certain relations for the elements of the sequence $\{F_i\}$.

\section{Universal probability laws}
Starting from the general form for the probability of $\{F_i\}$, Eq.~\eqref{p_seq_F}, we now ask simple questions on the probability of observing certain relations between the entries of the sequence $\{F_i\}$. As we shall show, there exist probabilities which are completely independent from the specific structure of the observable $f(\psi)$ and even, in certain cases, from the underlying dynamics, for example on the structure of the jump operators or the Hamiltonian. To make this more concrete already at this point, we anticipate here that the goal is to give an answer to questions like: ``What is the probability for the value $F_1$ to be larger than any other $F_i$?"  To this end, we will exploit the ideas put forward in Ref.~\cite{smith2023} for classical stochastic processes.

\subsection{Uncorrelated structure in Laplace space}

The starting point to answer questions like the one above consists in performing a Laplace transform of the probability in Eq.~\eqref{p_seq_F}, from the time-domain variable $t$ to the Laplace domain variable $s$. This reads as \cite{smith2023}
\begin{equation}
P(\{F_i\},N|s)=\int_0^\infty {\rm d }t \, e^{-s t} P(\{F_i\},N|t)\, . 
    \label{P_laplace}
\end{equation}
By exploiting the convolution structure of the probability $P(\{F_i\},N|t)$, we can write 
\begin{equation}
P(\{F_i\}, N|s)=Q(s)\Pi(F_{N+1},s)  q^N(s)\prod_{i=1}^{N}\pi (F_i,s)\, , 
\label{prob_Lapl}
\end{equation}
where we defined the ``normalized" Laplace transforms 
$$
\Pi(F,s)=\frac{1}{Q(s)}\int_0^\infty {\rm d}t  \, e^{-s t } {\rm Tr}\left( \tilde{\Lambda}^{F}_{t}[\psi_0]\right)\, ,
$$
as well as 
$$
\pi(F,s)=\frac{1}{q(s)}\int_0^\infty {\rm d}t  \, e^{-st } {\rm Tr}\left(\mathcal{W}\circ \tilde{\Lambda}^{F}_{t}[\psi_0]\right)\, .
$$
The corresponding normalizations are found by integrating the ``bare" Laplace transforms over all possible outcomes of $F$. That is, 
\begin{equation}
\begin{split}
    q(s):&=\int {\rm d}F \int_0^\infty {\rm d}t  \, e^{-s t } {\rm Tr}\left(\mathcal{W}\circ \tilde{\Lambda}^{F}_{t}[\psi_0]\right)\\
    &=\int_0^\infty {\rm d}t \, e^{-s t } {\rm Tr}\left(\mathcal{W}\circ e^{t(\mathcal{L}-\mathcal{R})}[\psi_0]\right)\, ,
    \end{split}
\end{equation}
and that 
\begin{equation}
\begin{split}
    Q(s):&=\int {\rm d}F \int_0^\infty {\rm d}t  \, e^{-s t } {\rm Tr}\left( \tilde{\Lambda}^{F}_{t}[\psi_0]\right)\\
    &=\int_0^\infty {\rm d}t \, e^{-s t } {\rm Tr}\left(e^{t(\mathcal{L}-\mathcal{R})}[\psi_0]\right)\, .
    \end{split}
\end{equation}
In both equations, we have exploited that 
\begin{equation}
\begin{split}
\int {\rm d}F \sum_{\forall \vec{t},\vec{\alpha}} \delta(F(\omega_{\vec{\tau},\vec{\alpha}}^{\psi_0})-{{F}})\tilde{\Lambda}_{\vec{t},\vec{\alpha}}[\psi_0]&=\sum_{\forall \vec{t},\vec{\alpha}}\tilde{\Lambda}_{\vec{t},\vec{\alpha}}[\psi_0]\\
&=e^{t(\mathcal{L}-\mathcal{R})}[\psi_0]\, .
\end{split}
\end{equation}
Similarly to the discussion related to Eq.~\eqref{normaliz}, the last equality in the above equation comes from the fact that the sum over all trajectories, free of reset processes, can be seen as a Dyson series expansion applied to the map $e^{t(\mathcal{L}-\mathcal{R})}$ and considering the map $\sum_{\alpha}\mathcal{J}_\alpha[\cdot]$ as an ``interaction" term. 

Interpreting $P(\{F_i\},N|s)$ as a probability, we see from Eq.~\eqref{prob_Lapl} that it has a product form which is furthermore symmetric under any exchange of the outcomes $F_i$ that do not involve the last term $F_{N+1}$. As discussed, this last value is indeed special since the time-window it refers to, unlike the others, does not end with a reset event. Only when considering a Poissonian reset process,  i.e., $\Gamma_m=\Gamma$ for all $m$, the probability $\Pi(F,s)$ is proportional to $\pi(F,s)$ which essentially ensures a fully permutation-symmetric character to $P(\{F_i\},N|s)$, as in the situations studied in Ref.~\cite{smith2023}. Therefore, for a generalized reset process, we integrate out the last random variable $F_{N+1}$. In such a case, we will only ask questions regarding the relative magnitude of the first $N$ random variables \cite{godreche2023}. Integrating out the last variable, we find 
\begin{equation}
    \label{P-prime-gen}
P'(\{F_i\},N|s)=Q(s)q^N(s)\prod_{i=1}^N \pi(F_i,s)\, . 
\end{equation}
This expression shows that the different $F_i$ are independent and identically distributed random variables.

\subsection{Probability of the maximum}
In the following we focus on a specific question concerning the sequence $\{F_i\}$: what is the probability for the first random variable $F_1$ to be the largest in the sequence? As shown in Ref.~\cite{smith2023}, the ideas that allow to answer this question extend to generic relations which solely probe the relative magnitude of the entries of the sequence. 

\subsubsection{Poissonian reset process}
\begin{figure}[t]
\centering
\includegraphics[width=\columnwidth]{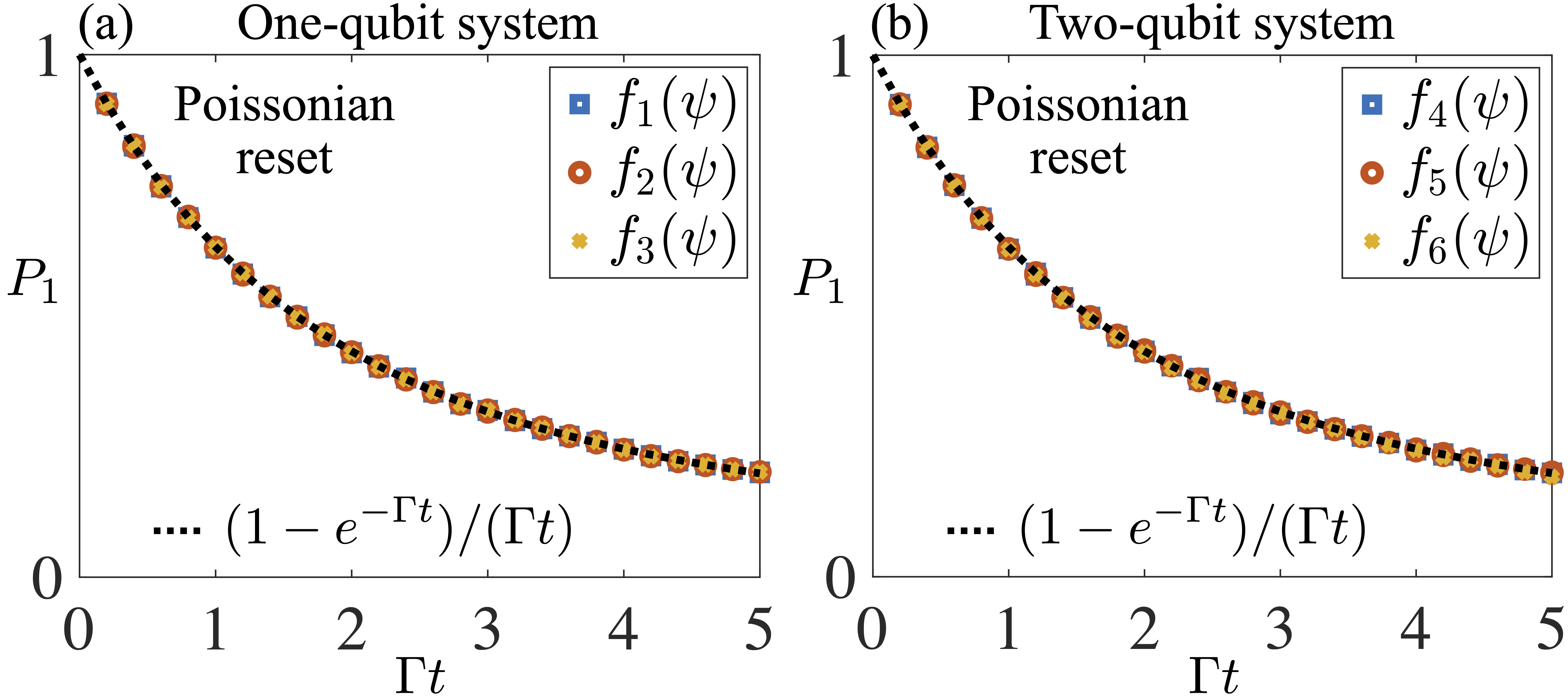}
\caption{{\bf Open quantum dynamics with Poissonian resets.} (a) Probability $P_1$  for the one-qubit system. Symbols refer to the different functions of the state discussed in Section \ref{systems}. We run 50000 trajectories for two different parameter regimes, $\Omega=\gamma_1=\Gamma$, $\gamma_2=2\Gamma$ and $\Omega=2\Gamma$, $\gamma_1=\Gamma$, $\gamma_2=\Gamma/2$, with $\Gamma$ being the Poissonian reset rate, $\Gamma_1=\Gamma_2=\Gamma$. (b) Probability $P_1$  for the two-qubit system. Symbols refer to the different functions of the state discussed in Section \ref{systems}. We run 50000 trajectories for two different parameter regimes, $\Omega=\gamma_1=\gamma_2=\Gamma$, $V=2\Gamma$, 
and $\Omega=\Gamma$, $V=\Gamma/2$, $\gamma_1=\gamma_2/2=\Gamma$, with $\Gamma$ being the Poissonian reset rate, $\Gamma_i=\Gamma$, for $i=1,2,3,4$. In both panels, numerical simulations are compared with the prediction shown after Eq.~\eqref{prediction_Poisson}, first derived in Ref.~\cite{smith2023}}
\label{Fig2}
\end{figure}

We start with the case of Poissonian resets, already discussed for classical processes in Ref.~\cite{smith2023}. 
The probability, $P_1(t)=\sum_N{\rm Prob}(F_1>F_i,i\in[2, N+1]|t)$, for $F_1$ to be the largest value in the sequence $\{F_i\}$, irrespective of the number of reset events, is given by 
\begin{equation}
\begin{split}
P_1(t)&=\sum_{N=0}^\infty \int \prod_{i=1}^{N+1}{\rm d}F_i \int \prod_{j=1}^{N} {\rm d}t_j \times \\
&\times P(\{F_i\},\{t_j\},N|t)\prod_{k=1}^{N+1}\theta(F_1-F_k)\, .
\end{split}
\end{equation}
The step function $\theta(F_1-F_k)$, which we define here as $\theta(x)=1$ for $x>0$ and $\theta(x)=0$ otherwise, implements the constraint that all $F_k$ are  smaller than $F_1$. 
We now calculate the Laplace transform of the above probability 
\begin{equation*}
    P_1(s)=\int_0^\infty {\rm d}t \, e^{-st} P_1(t)\, ,
\end{equation*}
and we find
\begin{equation}
P_1(s)=\sum_{N=0}^\infty \int \prod_{i=1}^{N+1}{\rm d}F_i P(\{F_i\},N|s)\prod_{k=1}^{N+1}\theta(F_1-F_k)\, .
\label{C1}
\end{equation}
For Poissonian resets, we have that $q(s)=\Gamma Q(s)$ while $\pi(F,s)=\Pi(F,s)$ so that the probability $P(\{F_i\},N|s)$ is completely invariant under any permutation of the $F_i$. As such, the probability that $F_1$ is larger than the other random variables is equal to the probability that any other $F_i$ being larger than the others. This suggests that the constraint can be actually removed and substituted by a factor $(N+1)^{-1}$. Overall, this yields 
$$
P_1(s)=\frac{1}{\Gamma} \sum_{N=0}^\infty \frac{q^{N+1}(s)}{N+1}\, ,
$$
which, considering that for a Poisson reset process $q(s)=\Gamma/(\Gamma+s)$, leads to 
\begin{equation}
P_1(s)=\frac{1}{\Gamma}\log \frac{s+\Gamma}{\Gamma}\, ,
\label{prediction_Poisson}
\end{equation}
as derived in Ref.~\cite{smith2023}. Moving back to the time-domain by applying the inverse Laplace transform to $P_1(s)$ one obtains 
$P_1(t)=(1-e^{-\Gamma t})/(\Gamma t)$. The universal character of this probability is  confirmed in Fig.~\ref{Fig2}(a-b), considering two different quantum systems [see details in Sec.~\ref{systems}] for which we study two different parameter regimes for each system and different observables. 

\subsubsection{Generalized reset process}
For a generalized reset process, one has that $\pi(F,s)$ is not equal nor proportional to $\Pi(F,s)$. Recalling Eq.~\eqref{prob_Lapl}, this means that the probability $P(\{F_i\},N|s)$ is only invariant when considering permutations of the first $N$ entries of the sequence $\{F_i\}_i$. We thus only consider relations involving the first $N$ random variables $F_i$. This means that we can integrate out $F_{N+1}$ and work with the probability $P'(\{F_i\},N|s)$ defined in Eq.~\eqref{P-prime-gen}, as also done in Ref.~\cite{godreche2023} in a classical setup. 

\begin{figure}[t]
\centering
\includegraphics[width=\columnwidth]{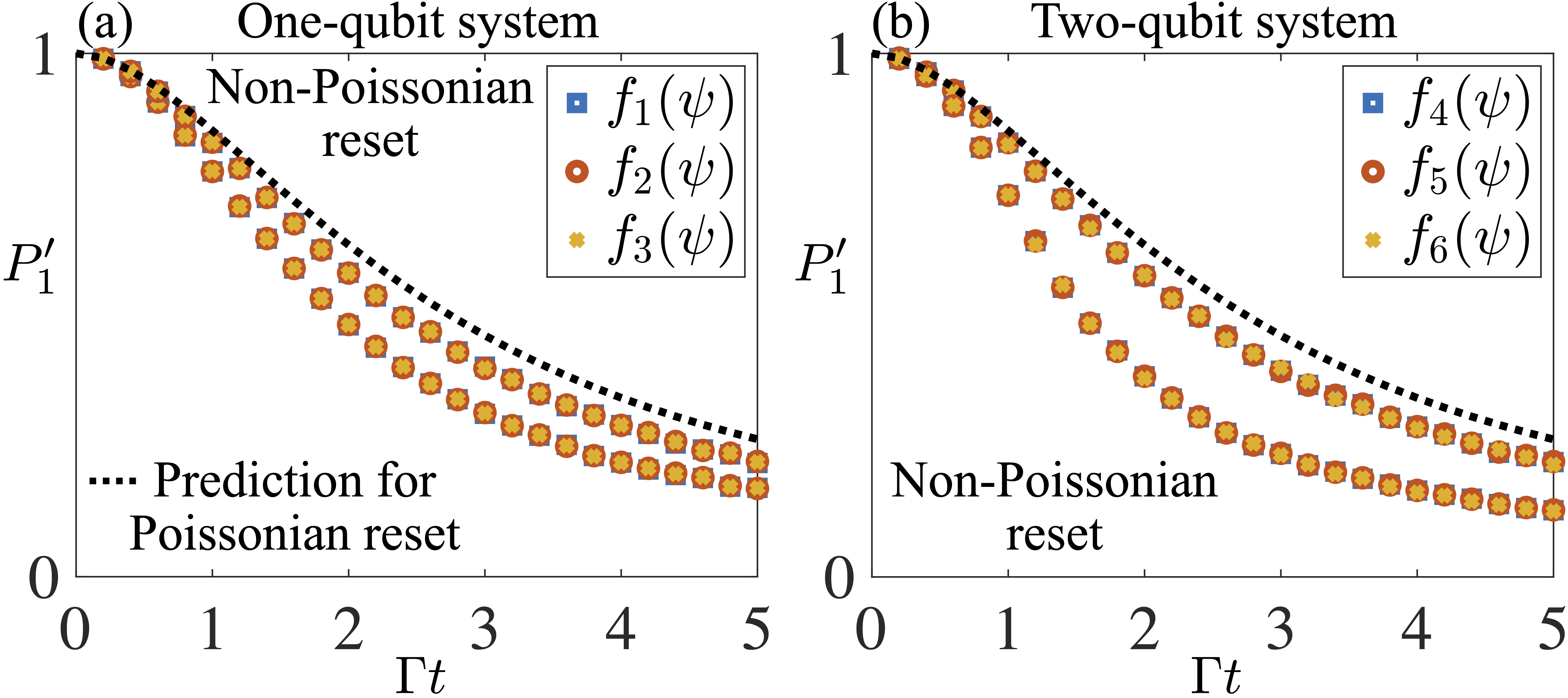}
\caption{{\bf Open quantum dynamics with non-Poissonian resets.} (a) Probability $P_1'$  for the one-qubit system. Symbols refer to the different functions of the state discussed in Section \ref{systems}. We run 50000 trajectories for two different parameter regimes, $\Omega=\gamma_1=\Gamma$, $\gamma_2=2\Gamma$, and $\Omega=\Gamma/10$, $\gamma_1=\gamma_2/2=\Gamma$, both with reset rates $\Gamma_1=\Gamma_2/2=\Gamma$. (b) Probability $P_1'$  for the two-qubit system. Symbols refer to the different functions of the state discussed in Section \ref{systems}. We run 50000 trajectories for two different parameter regimes, $\Omega=V/2=\Gamma$ $\gamma_1=\gamma_2=\Gamma$, and $\Omega=2\Gamma$, $V=\Gamma/4$, $\gamma_1=\gamma_2=1$. The state-dependent reset rates are, in both cases, $\Gamma_1=\Gamma$, $\Gamma_2=\Gamma_3=2\Gamma$, $\Gamma_4=4\Gamma$. In both panels, as a reference, we provide the  prediction in Eq.~\eqref{prediction_P1'_Poisson} for a Poissonian reset with rate $\Gamma$.}
\label{Fig3}
\end{figure}

The probability, $P_1'(t)=\sum_N{\rm Prob}(A_1>A_i,i\in[2, N]|t)$, for $F_1$ to be the largest value among the first $N$ entries of the  sequence $\{F_i\}$, irrespective of the number of reset events, is given in Laplace domain by the relation 
\begin{equation}
\begin{split}
P_1'(s)&=\sum_{N=0}^\infty \int \prod_{i=1}^{N}{\rm d}F_i P'(\{F_i\},N|s)\prod_{k=1}^{N}\theta(F_1-F_k)\, .
\label{C2}
\end{split}
\end{equation}
Analogously to the previous discussion, the constraint can be substituted by the factor $1/N$. Integrating over the random variables we then find 
\begin{equation*}
P_1'(s)=Q(s)\left[1+\sum_{N=1}^\infty\frac{q^N(s)}{N}\right]=Q(s)-Q(s)\log[1-q(s)]\, .
\end{equation*}
We note that this probability also includes the case in which there are no reset events and the case in which there is only one reset event. Moreover, we note that the above probability does not depend on the specific structure of the chosen function $f(\psi)$. However, it generically depends on the precise structure of the open quantum dynamics. This can be understood by inspecting the definition of the functions $Q(s)$ and $q(s)$, which are determined by the interplay  between the generator $\mathcal{L}$, the no-reset  dynamics $\mathcal{R}$ and the reset map $\mathcal{W}$. The dependence on $\mathcal{L}$ disappears in the case of Poissonian reset processes for which $\mathcal{R}$ is proportional to the identity map while $\mathcal{W}$ essentially acts as the trace operation. Importantly, this means that in the situation of non-Poissonian resets the probability is universal solely in the sense that it does not depend on the considered observable but it is instead sensitive to the open quantum dynamics of the problem. This observation is supported by numerical results shown in Fig.~\ref{Fig3}(a-b). 

The probability $P_1'(t)$ can in principle be calculated by applying the inverse Laplace transform to $P'_1(s)$, even though exact analytical expression may be difficult to get.  Numerical results for $P_1'(t)$, obtained by simulating quantum trajectories for two exemplaric stochastic processes, are shown in Fig.~\ref{Fig3}(a-b). In the case of Poissonian resets, the probability $P_1'(t)$ can also be computed analytically \cite{godreche2023}
\begin{equation}
P_1'(t)=e^{-\Gamma t}\left(1-\gamma_{\rm EM} -\ln \Gamma t+\int_{-\infty}^{\Gamma t} {\rm d} x\frac{e^{x}}{x}\right)\, , 
\label{prediction_P1'_Poisson}
\end{equation}
where $\gamma_{\rm EM}\approx 0.5772$ is the Euler–Mascheroni constant.

\subsection{Numerical benchmarks}
\label{systems}
To verify numerically our findings, we consider two simple open quantum systems, which we will introduce in the following.
\subsubsection{One-qubit system}
The first is a single qubit, for which we chose the basis states $\ket{\uparrow},\ket{\downarrow}$ and which evolves under the Hamiltonian $H=\Omega \sigma_x$, with $\sigma_x=\ket{\uparrow}\!\bra{\downarrow}+\ket{\downarrow}\!\bra{\uparrow}$. Dissipation is governed by the two jump operators $J_1=\ket{\downarrow}\!\bra{\uparrow}$ and $J_2=\ket{\uparrow}\!\bra{\downarrow}$, which are associated with rates $\gamma_1$ and  $\gamma_2$, respectively. The reset map [cf.~Eq.~\eqref{reset_jump_operators}] is instead
$$
\mathcal{W}[\rho]=\Gamma_1 \ket{\downarrow}\!\bra{\downarrow}\rho \ket{\downarrow}\!\bra{\downarrow}+ \Gamma_2 \ket{\downarrow}\!\bra{\uparrow}\rho \ket{\uparrow}\!\bra{\downarrow}\, .
$$
The latter accounts for a process that resets the system to state $\ket{\downarrow}$ with rate $\Gamma_1$ if the system is in $\ket{\downarrow}$ and with rate $\Gamma_2$ if the system is in $\ket{\uparrow}$. This also defines the map $\mathcal{R}$. For this system, we consider  observables of quantum trajectories obtained by considering the following functions, see definition in  Eq.~\eqref{time_int_state}, 
$$
f_1(\psi)=\left|\bra{\downarrow}\psi\ket{\uparrow}\right|\, , \qquad  f_2(\psi)=\bra{\uparrow}\psi\ket{\uparrow}\, ,
$$
as well as $f_3=f_1+f_2$. The function $f_1$ quantifies quantum superposition between states $\ket{\uparrow}$ and $\ket{\downarrow}$, the function $f_2$ encodes the probability of finding  the system in $\ket{\uparrow}$,   while $f_3$ is an arbitrarily chosen combination of the two.

\subsubsection{Two-qubit system}
The second example system that we consider  is a two-qubit quantum system with the Ising-model Hamiltonian 
$$
H=\Omega(\sigma_x^{(1)}+\sigma_x^{(2)})+V\sigma_z^{(1)}\sigma_z^{(2)}\, , 
$$ 
where the superscript on the operators indicates the qubit  the latter are referring to. As for the other system, $\sigma_x=\ket{\uparrow}\!\bra{\downarrow}+\ket{\downarrow}\!\bra{\uparrow}$ while $\sigma_z=\ket{\uparrow}\!\bra{\uparrow}-\ket{\downarrow}\!\bra{\downarrow}$. As jump operators, we consider the ladder operator $\sigma_-=\ket{\downarrow}\!\bra{\uparrow}$ for each particle, i.e., 
$$
J_1=\sigma_-^{(1)}\, , \qquad J_2=\sigma_-^{(2)}\, ,
$$
associated with the rates $\gamma_1$ and $\gamma_2$, respectively. 
For such a system, the reset process is constructed such that the four possible basis states are associated with different reset rates as follows
\begin{equation}
\begin{split}
\ket{\downarrow\downarrow}\leftrightarrow\Gamma_1\, ,\qquad \ket{\uparrow\downarrow}\leftrightarrow\Gamma_2\, ,\\
\ket{\downarrow\uparrow}\leftrightarrow\Gamma_3\, ,\qquad \ket{\uparrow\uparrow}\leftrightarrow\Gamma_4\, .
\end{split}
\end{equation}
In this case, we chose the state $\ket{\downarrow\downarrow}$ as the reset state. 

As a first observable for this second system, we consider a measure of the entanglement content of the single dynamical realizations of the stochastic process. Entanglement in quantum trajectories is receiving much attention nowadays due to the recent interest in its dynamics in many-body systems, in the emergence of measurement-induced phase transitions, and in the study of the complexity of the numerical simulation of open quantum systems (see for example Refs.~\cite{nahum2017,nahum2018,keyserlingk2018,li2018,skinner2019,li2019,chan2019,gullans2020,alberton2021,vovk2022}). 
Since we consider an initial pure state, we can quantify  entanglement for  our two-qubit system via the von Neumann entanglement entropy, which we calculate through the reduced state $\rho_1(\psi)={\rm Tr}_2 \left(\psi\right)$ as 
$$
f_4(\psi)=-{\rm Tr}_1 \left[\rho_1(\psi)\ln \rho_1(\psi)\right]\, .
$$
Here, ${\rm Tr}_i$ denotes the trace over the $i$th
qubit. As a further observable, we calculate the number of qubits in state $\ket{\uparrow}$ as
$$
f_5(\psi)={\rm Tr}\left[\psi (n^{(1)}+n^{(2)})\right]\, ,
$$
where we defined $n=\ket{\uparrow}\!\bra{\uparrow}$, which is related to the global magnetization of the considered open quantum  Ising model. As a last function, we arbitrarily consider the difference between the previous two, $f_6(\psi)=f_4(\psi)-f_5(\psi)$. 

\section{Emergent universal probabilities for jump-related observables }
In this Section, we consider trajectory observables which do not directly depend on the state of the system, but which are rather  defined through the total counts of the jump events that occur during the open system dynamics. 

Let us assume that the dynamics in Eq.~\eqref{eq:stoc} is observed for a total time $t$. During this time-interval, the system will undergo jumps associated with the operator $J_\alpha$, a total of $K_\alpha$ times where 
$$
K_\alpha =\int_0^t {\rm d}n_u^\alpha\, .
$$
Clearly, the quantities $K_\alpha$, for $\alpha=1,2,\dots D$, where $D$ is the total number of jump operators [cf.~Eq.~\eqref{lindblad}], are stochastic due to the random nature of the noises ${\rm d}n_t^\alpha$.
Introducing the vector ${K}=(K_1,K_2,\dots K_D)$, it is natural to ask what is the probability of observing a trajectory with exactly $K=\bar{K}$. To construct this probability, it is convenient to first note that the probability $P(K=0|t)$, for all $K_\alpha$ to be zero up to time $t$, is given by 
\begin{equation}
P(K=0|t)={\rm Tr}\left(\Phi^{K=0}_t[\psi_0]\right)\, , \mbox{ with }\,  \Phi^{K=0}_t[\cdot]:=e^{t\mathcal{L}_\infty}[\cdot]\, .
\label{Phi_k}
\end{equation}
The above result is obtained by considering that the probability for not jumping over a single time-step is approximately ${\rm Tr}\left(e^{{\rm d} u \mathcal{L}_\infty}[\psi_u]\right)$, and that we have to take products of these quantities recalling the no-jump evolution in Eq.~\eqref{eq:stoc}. We can now recursively define the maps 
\begin{equation}
\label{Prob_K_iteration}
\Phi^K_t[\cdot]=\sum_{\alpha=1}^D \int_0^t {\rm d}u \, \Phi_{t-u}^{0}\circ\mathcal{J}_\alpha\circ \Phi^{K-{\rm e}_\alpha}_u[\cdot]\, ,
\end{equation}
where we have chosen the basis vector ${\rm e}_\alpha$ to be the vector of elements equal to zero apart from the $\alpha$th element, which is set to one. With these maps one obtains the desired probability as $P(K=\bar{K}|t)={\rm Tr}\left(\Phi_t^{\bar K}[\psi_0]\right)$. The idea behind this is that one can obtain an outcome $\bar{K}$ at time $t$ by having $K=\bar{K}-{\rm e}_\alpha$, for some $\alpha$, up to a time $u\le t$, followed by a jump of type $\alpha$ and a no-jump evolution from $u$ to $t$. This sequence of events corresponds to the composition of the maps inside the integral on the right hand side of of Eq.~\eqref{Prob_K_iteration}. Considering these occurrences for any $\alpha$ and for any $0\le u\le t$ thus gives the probability $P(K=\bar{K}|t)={\rm Tr}\left(\Phi_t^{\bar K}[\psi_0]\right)$. 

From the latter probabilities $P(K|t)$, we can construct the probability of observing an outcome $\bar{A}$ for any possible function $A$ of the (activity) vector $K$, as 
$$
P(A=\bar{A}|t)=\sum_{\forall K}\delta_{\bar{A},g(K)}P(K|t)\, ,
$$
where we have assumed that $A=g(K)$.
Possible observables $A$ are for instance the total activity $A=\sum_\alpha K_\alpha$ or the imbalance between two different total counts $A=K_\alpha-K_\beta$. 

As also illustrated in Fig.~\ref{Fig1}(b), in the presence of the reset process the quantum trajectory is partitioned in different time-intervals separated by reset events. In each time-interval $i$, one has a value $A_i$ for the chosen observable [cf.~Fig.~\ref{Fig1}(b)]. Following the steps leading to Eq.~\eqref{p_seq_F}, we can write the probability of observing $N$ reset events and a sequence $\{A_i\}=\{A_1,A_2,\dots A_{N+1}\}$ for the observable $A$ as 
\begin{equation}
\begin{split}
P(\{A_i\},N|t)&=\int_0^t {\rm d}t_{N}\int_0^{t_{N}}{\rm d }t_{N-1}\dots \int_0^{t_2}{\rm d}t_1 \times \\
&\times {\rm Tr}\left( \tilde{\Psi}^{A_{N+1}}_{\tau_{N+1}}[\psi_0]\right) \prod_{i=1}^{N}{\rm Tr}\left(\mathcal{W}\circ\tilde{\Psi}^{A_{i}}_{\tau_i}[\psi_0]\right)\, .
\label{p_seq_A}
\end{split}
\end{equation}
Here, we defined the maps 
$$
\tilde{\Psi}^{\bar{A}}_t[\cdot]=\sum_{\forall K}\delta_{\bar{A},g(K)} \tilde{\Phi}^K_t[\cdot ]\, ,
$$
where $\tilde{\Phi}_t^K$ are the analogous of the map ${\Phi}_t^K$ introduced in Eq.~\eqref{Phi_k} but with $\tilde{\Phi}_t^0=e^{t\tilde{\mathcal{L}}_\infty}$ which accounts also for the presence of the no-reset dynamics.

\begin{figure}[t]
\centering
\includegraphics[width=\columnwidth]{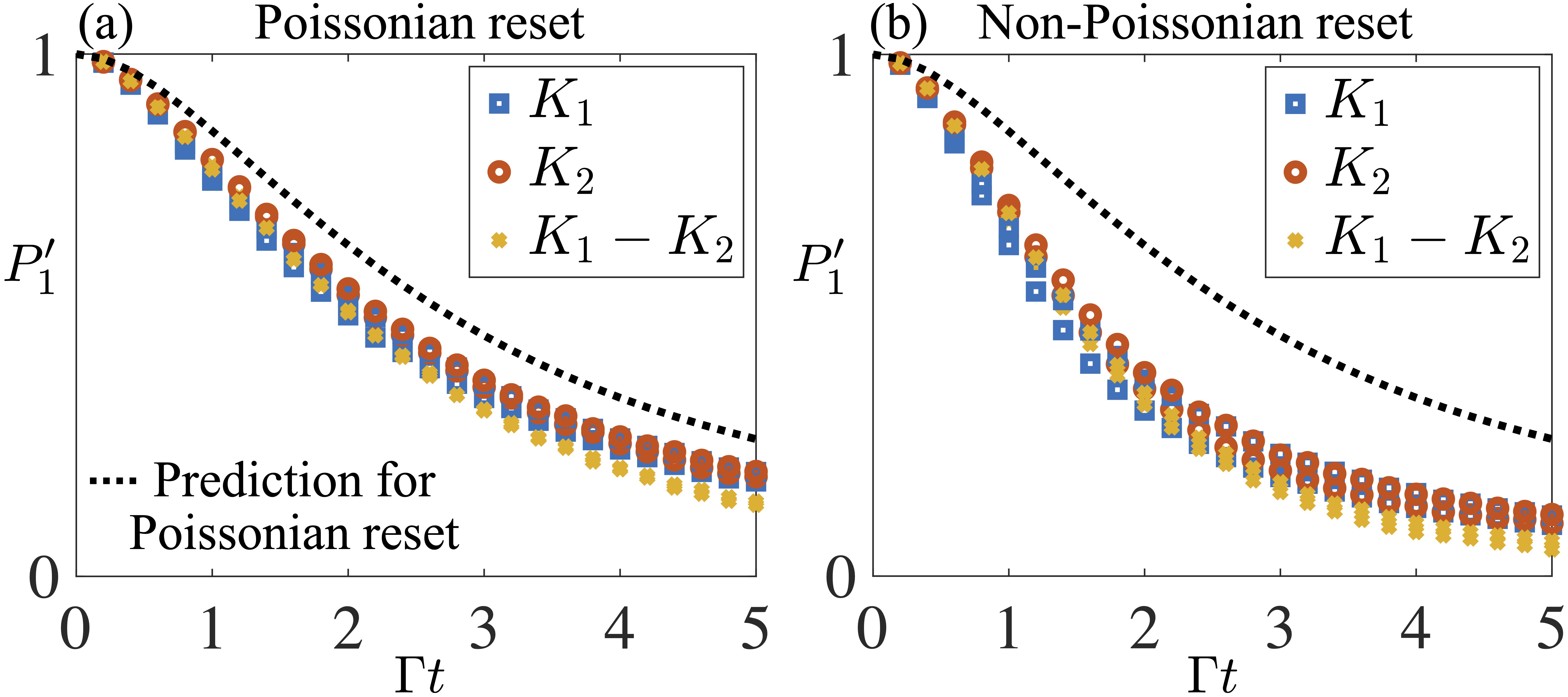}
\caption{{\bf One-qubit system: quantum-jump  observables.} (a) Probability $P_1'$  for the one-qubit system. Symbols refer to different functions of the quantum-jump activities. We run 50000 trajectories for two different parameter regimes, $\Omega=\gamma_1=\gamma_2/2=\Gamma$ and $\Omega=\gamma_1/2=\gamma_2/2=\Gamma$, where $\Gamma$ is the Poissonian reset rate. (b) Probability $P_1'$  for the one-qubit system. Symbols refer to the different functions of the quantum-jump activities. We run 50000 trajectories for two different parameter regimes, $\Omega=\gamma_1=\gamma_2/2=\Gamma$ and $\Omega=\gamma_1/2=\gamma_2/2=\Gamma$, with reset rates $\Gamma_1=\Gamma$ and $\Gamma_2=2\Gamma$. In both panels, as a reference, we provide the  prediction in Eq.~\eqref{prediction_P1'_Poisson} for a Poissonian reset with rate $\Gamma$.}
\label{Fig4}
\end{figure} 

We now ask what is the probability $P_1'(t)=\sum_N{\rm Prob}(A_1>A_i,i\in[2, N]|t)$, for $A_1$ to be the largest between the first $N$ entries of  the sequence $\{A_i\}$, irrespective of the number of reset events. In the Laplace domain, this probability is given by 
\begin{equation}
\begin{split}
P_1'(s)=\sum_{N=0}^\infty \int \prod_{i=1}^{N}{\rm d}A_i P'(\{A_i\},N|s)\prod_{k=1}^{N}\theta(A_1-A_k)\, ,
\end{split}
\label{prob_discrete}
\end{equation}
where we have defined $P'(\{A_i\},N|s)=\int {\rm d}A_{N+1} P(\{A_i\},N|s)$. Inspecting the above structure, one would be tempted to say that, also in this case, due to permutation invariance, the constraint implemented by the step functions can be substituted by a factor $1/N$. This would then imply that, i) for a Poissonian reset process this probability does not depend on the chosen observable nor on the specific details of the dynamics \cite{smith2023} and that ii) for a non-Poissonian reset process the probability is independent on the chosen observable.  However, by looking at Fig.~\ref{Fig4}(a-b), we see that this is actually not correct. For both the Poissonian and the non-Poissonian case, the probability is not universal and depends on the details of the dynamics as well as on the chosen observable. 

The reason for this difference is the following (see also discussion in the next Section). The observable $A$ is a discrete observable so that there is also a nonzero probability for observing the same outcome for different entries of the sequence $\{A_i\}$, which is generically not possible for the trajectory observables defined in Eq.~\eqref{time_int_state}. This means that, in a random  sequence, it can also happen that $A_1=A_i$, for some $i$. As it will become evident through the example discussed in Section
\ref{min_model}, this fact makes the 
the probability of Eq.~\eqref{prob_discrete} dependent on  the considered observable. Essentially, this is due to the fact that the probability of finding two, or more, equal outcomes in a sequence depends on the probability distribution of the observable in a single time-interval.

\begin{figure}[t]
\centering
\includegraphics[width=\columnwidth]{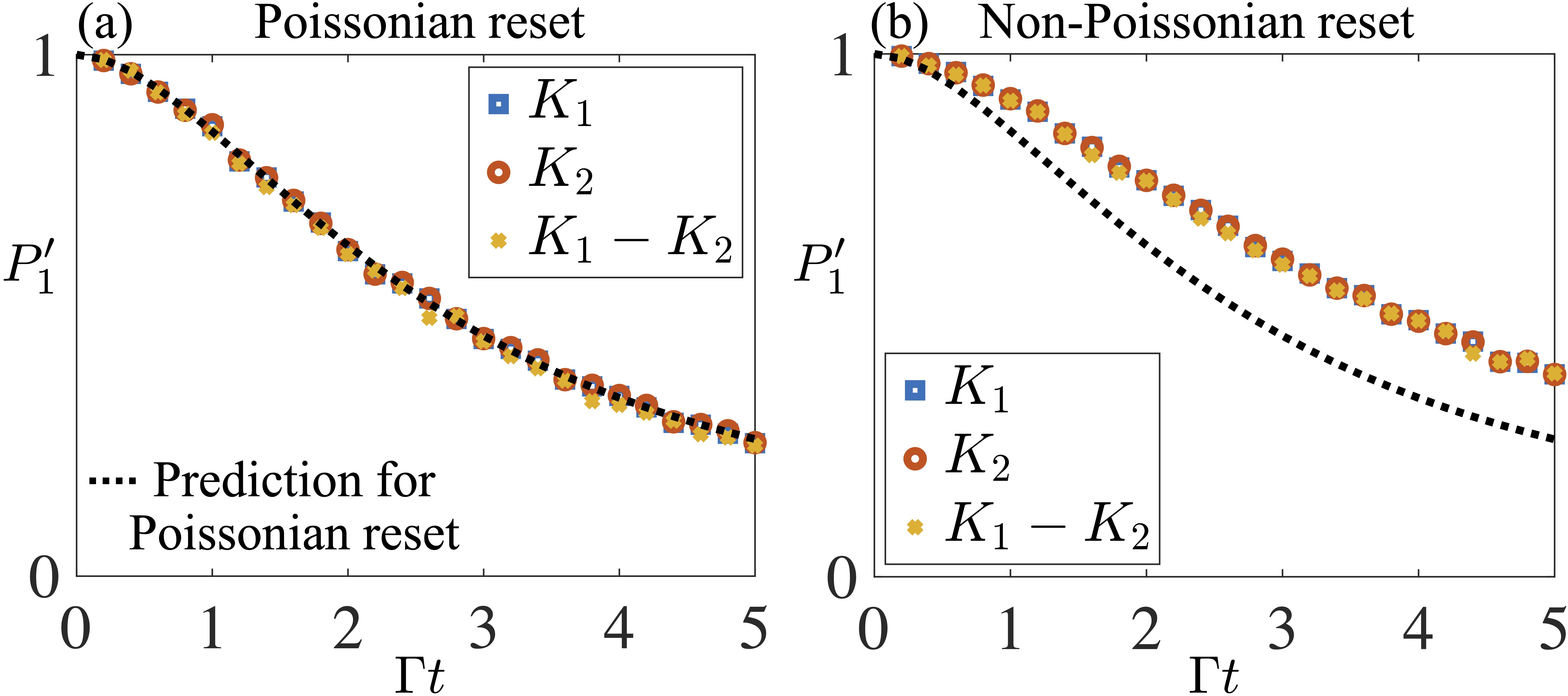}
\caption{{\bf Emergent universality for quantum-jump observables in the weak reset-rate limit.} (a) Probability $P_1'$  for the one-qubit system. Symbols refer to different functions of the quantum-jump activities. We run 5000 trajectories, for which fluctuations are evident, for a parameter regime with  $\Omega=\gamma_1=\gamma_2/2=100\Gamma$, where $\Gamma$ is the (weak) Poissonian reset rate. (b) Same as in (a) but for a non-Poissonian reset process specified by the rates $\Gamma_1=2\Gamma_2=\Omega/100$. In both panels, as a reference, we provide the  prediction in Eq.~\eqref{prediction_P1'_Poisson} for a Poissonian reset with rate $\Gamma$. }
\label{Fig5}
\end{figure} 

For a discrete random variable a nonzero probability of having two equal outcomes in the sequence can only be avoided if each  outcome of the random variable occurs with a vanishingly small probability (see also discussion in Section \ref{min_model}). The way to achieve this in our setting is by making  the time-intervals in between reset events extremely long. In this way, even though jump events can still only assume discrete values, the number of possible outcomes increases. Indeed, since jump observables obey a law of large numbers with time being the large parameter, for long time-intervals $\tau$ the value of all $K_\alpha$'s will become extensive in $\tau$ as well as their variance, due to central limit theorems. In essence, the rates $K_\alpha/\tau$ tend to become continuous variables, which means that the probability of observing a specific value of $K_\alpha$ becomes small. Long time-intervals between reset events can be achieved by choosing vanishingly small reset rates. In Fig.~\ref{Fig5}(a-b), we show indeed that in these cases, universal probability laws are recovered both for Poissonian and non-Poissonian processes (we only present results for the single-qubit system). 

\section{A minimal model}
\label{min_model}
In this last Section, we rationalize the findings of this paper, constructing the simplest possible model which still displays the essential features of our observations. As mentioned in the Introduction, the presence of the reset process makes the different random variables, which we have considered,  statistically independent. In order to better appreciate the role of the probability distribution of the random variables themselves, we consider here the case of two identically distributed and statistically independent random variables. This amount to consider a sequence with only two elements. We call these variables $x_1$ and $x_2$ and we are concerned with characterizing the probability $P(x_1>x_2)$ of observing that $x_1>x_2$. 

\subsection{Continuous random variables}
We begin by considering the case of variables assuming values from a continuous set. We assume such set to be the set of real numbers and characterize these random variables through their  probability density function $\mu(x)$. 
The probability that $x_1>x_2$ (i.e., that $x_1$ is the largest value in the simple sequence considered) can thus be written as 
$$
P(x_1>x_2)=\int_{-\infty}^\infty {\rm d}x_1  \mu(x_1)\int_{-\infty }^{x_1}{\rm d}x_2 \mu(x_2) \, .
$$
The integral over the probability density function gives the cumulative probability function $C(x)=\int_{-\infty}^{x}{\rm d}x_2 \, \mu(x_2)$ so that we have
$$
P(x_1>x_2)=\int_{-\infty}^\infty {\rm d}x_1  \mu(x_1) C(x_1) \, .
$$
Recognizing that $\mu(x)={\rm d}C(x)/{\rm d}x$ and changing variables in the integration we have 
$$
P(x_1>x_2)=\int_{0}^1 {\rm d}C \, C =\frac{1}{2}.
$$
This result reflects the intuitive observation that there is no reason to expect a different probability for $x_1>x_2$ or $x_2>x_1$ given that the two random variables are statistically independent. This is essentially what allows one to substitute to the constraints in Eq.~\eqref{C1} and Eq.~\eqref{C2} the inverse of the number of possible outcomes (here given by $1/2$). 

\subsection{Discrete random variable}
We now consider the case in which the two variables can assume the values $k=0,1,2, \dots M$ with probability $P(x=k)=p_k$. The probability that $x_1>x_2$ can now be written as 
$$
P(x_1>x_2)=\sum_{\forall k,k':k>k'}p_kp_{k'}\, .
$$
Clearly, also in this case there is no reason why one should expect $P(x_1>x_2)\neq P(x_2<x_1)$, so that in fact we can write  
\begin{equation}
\begin{split}
P(x_1>x_2)&=\frac{P(x_1>x_2)+P(x_1<x_2)}{2}\\
&=\frac{1}{2}\left(1-\sum_{k=0}^Mp_k^2\right)\, .
\end{split}
\label{dis}
\end{equation}
The second equality comes from the fact that the probability that $x_1>x_2$ or $x_2>x_1$ is equal to one minus the probability that the two outcomes are equal. This shows that in this case, $P(x_1>x_2)$ is not given by the ``universal" value $1/2$, but actually depends on the details of the random variable and its probability distribution. 

\subsection{Emergent universal probability}
We now proceed with the previous example and show how the ``universal" value $1/2$ can emerge in a given limit. 
Let us now assume that $x_1,x_2$ are the total number of successes in a Bernoulli process made by $M$ repetition of a binary event. This means that 
$$
p_k=\binom{M}{k} p^k(1-p)^{M-k}\, ,
$$
where $p$ is the probability of success for the single binary event. For any finite $M$, we have that 
the probability of finding two equal values for $x_1$ and $x_2$ remains finite. However, when letting $M\to\infty$ all the single probabilities $p_k$ become vanishingly small and such that 
$\sum_{k=0}^Mp_k^2\to 0$, for $M\to\infty$. In this regime, one can therefore asymptotically recover the universal behavior $P(x_1>x_2)\to 1/2$. 

This simple example may seem unrelated to the quantum processes discussed above. However, in  a simplified picture, one could think of $p$ as being the probability of observing an emission event [cf.~Eq.~\eqref{p_event}], during the infinitesimal time-step ${\rm d}u$. The number of repetition $M$ can be associated with the number of (discrete) updates $\tau/{\rm d}u$, so that the larger $M$, the larger the resulting time-interval $\tau$. Here, thus, the large $M$ limit encodes the weak reset-rate limit discussed above. Essentially, one difference compared to the quantum process is that we only considered here a binary event which would corresponds to a single jump operator. Another one is that, in the quantum process, the probability $p$ generically depends on time, i.e., on the specific repetition performed and on the whole history of the previous outcomes.

\section{Discussion}

We have investigated the emergence of universal probability laws in quantum stochastic processes undergoing reset events.  Following the ideas first put forward in Ref.~\cite{smith2023} for classical reset processes, we have demonstrated that the probability of observing a given relation between the entries of a sequence of random variables, defined by the presence of reset events and depending on the quantum process, is universal if the considered variable assumes values in a  continuous set \cite{godreche2023}. In particular, for Poissonian reset this probability does not depend either on the dynamics or on the specific form of the chosen observable. For non-Poissonian resets, the probability depends instead on the details of the open quantum dynamics under consideration. When the random variables assume discrete values, the probability loses its universal character. This is essentially due to the fact that there is a nonzero probability of observing two, or more, equal outcomes in the sequence. Emergent universal probabilities still emerge, in these cases, when considering weak reset rates. 

Our findings generalize to classical Markov processes, which can also be formulated within the Lindblad formalism. In the present work, we illustrated our ideas using a specific probability, namely that of the first element of a sequence of trajectory observables being larger than all the others. 
However, our results generalize to the probability of other events involving possible orderings between the elements of the random sequence identified by the reset events as in Ref.~\cite{smith2023}.

\section*{Acknowledgments}
We acknowledge funding from the Deutsche Forschungsgemeinschaft (DFG, German Research Foundation) under Project No. 435696605 and through the Research Unit FOR 5413/1, Grant No. 465199066 as well as the Research Unit FOR 5522/1, Grant No. 499180199. We are also grateful for funding from the European Union’s Horizon Europe research and innovation program under Grant Agreement No. 101046968 (BRISQ). FC is indebted to the Baden-W\"urttemberg Stiftung for the financial support by the Eliteprogramme for Postdocs. We acknowledge support from EPSRC Grant No. EP/V031201/1 (IL and JPG).

\bibliography{refs}
\end{document}